\documentclass[aps,prl,twocolumn,amsfonts,amssymb,amsmath,floatfix,10pt,showpacs]{revtex4-1}
\usepackage[latin1]{inputenc}
\usepackage[T1]{fontenc}
\usepackage[english]{babel}
\usepackage{graphicx}
\usepackage{ulem}
\usepackage{hyperref}
\usepackage{siunitx}

\begin{document}

\title{Multi-band spectroscopy of ultracold fermions: Observation of reduced tunneling in attractive Bose-Fermi mixtures}

\author{J.~Heinze}

\author{S.~G\"otze}

\author{J.~S.~Krauser}

\author{B.~Hundt}

\author{N.~Fl\"aschner}

\author{D.-S.~L\"uhmann}

\author{C.~Becker}

\author{K.~Sengstock}

\affiliation{Institut f\"ur Laser-Physik, Universit\"at Hamburg, Luruper Chaussee 149, 22761 Hamburg, Germany}

\date{\today}

\begin{abstract}
\pacs{37.10.Jk, 67.85.Pq, 71.10.Fd, 78.47.-p}
\keywords{Ultracold quantum gases; Optical lattices; Bose-Fermi mixtures; Hubbard model; Momentum-resolved spectroscopy;}
\preprint{<report number>}

We perform a detailed experimental study of the band excitations and tunneling properties of ultracold fermions in optical lattices. Employing a novel multi-band spectroscopy for fermionic atoms we can measure the full band structure and tunneling energy with high accuracy. In an attractive Bose-Fermi mixture we observe a significant reduction of the fermionic tunneling energy, which depends on the relative atom numbers. We attribute this to an interaction-induced increase of the lattice depth due to self-trapping of the atoms.

\end{abstract}

\maketitle

Quantum gases in optical lattices have proven to be a versatile model system to investigate unsolved many-body problems from condensed-matter theories in a highly controlled fashion \cite{Bloch2008}. The tunability of interaction strength, lattice depth, and lattice geometry makes these systems particularly well suited to explore dependencies and limits of anticipated models like the Hubbard Hamiltonian \cite{Jaksch1998,Greiner2002}. Especially fermionic quantum gases have attracted much interest over the last years due to their close analogy to electrons in crystals. This led, e.g., to the observation of Fermi-surfaces \cite{Koehl2005}, antibunching \cite{Rom2006}, and the fermionic Mott insulator \cite{Joerdens2008,Schneider2008}. At the same time quantum gases in optical lattices allow to realize completely new systems like mixtures of bosonic and fermionic atoms where several exotic phases have been predicted. Among them are density waves \cite{Lewenstein2004}, supersolids \cite{Titvinidze2009}, and polaronic quasiparticles \cite{Mathey2004}. Pioneering experiments found a pronounced shift of the bosonic Mott insulator transition \cite{Ospelkaus2006a,Guenter2006,Best2008} and more recently a renormalization of the on-site interaction \cite{Will2010}. The realization of these systems over a broad range of parameters has extended the theoretical interest. There is a controversial debate on the correct theoretical description of Bose-Fermi mixtures: Various models have been discussed, including adiabatic heating during the lattice loading \cite{Cramer2008} and an effective potential approach including the admixture of higher bands \cite{Best2008,Luehmann2008selftrapping}. A recent ansatz introduces an extended Hubbard model with nearest neighbor interaction \cite{Mering2011}. For a complete comparison with experiments, it is important to independently measure both the effective on-site interaction and the effective tunneling in the mixture, where the latter has not been directly observed in experiments yet.

\begin{figure}[b]
  \centering
  \includegraphics[width=\columnwidth]{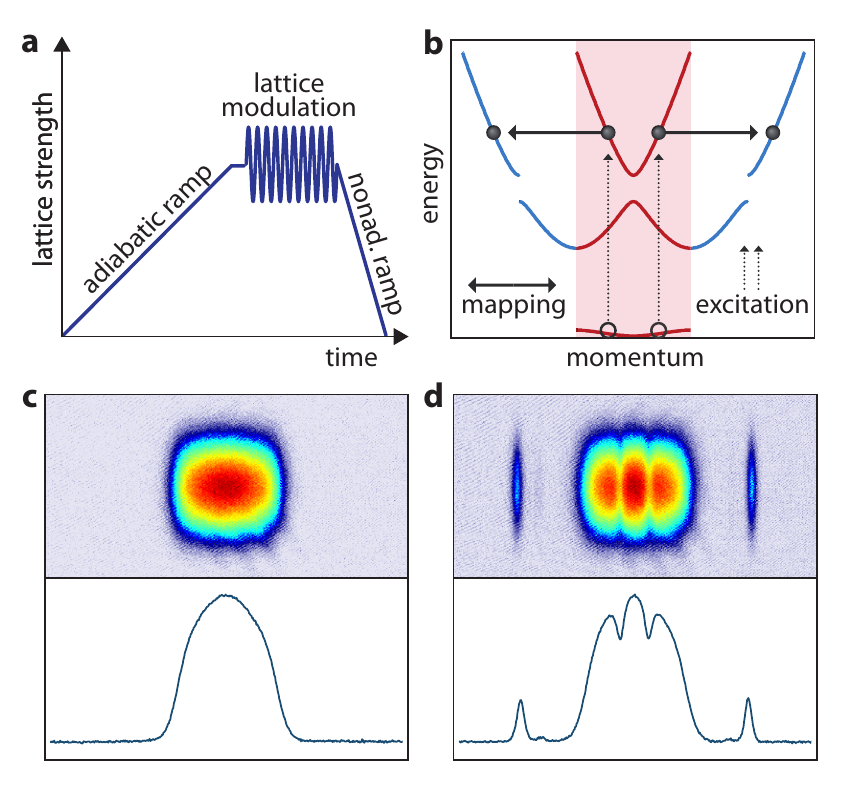}
  \caption{(color online). (a) Experimental sequence. (b) Experimental procedure in momentum space. (c) Typical absorption image and column sum with nonresonant modulation. The atoms occupy the first Brillouin zone. (d) With resonant modulation: Particle-hole excitations are clearly visible.}
  \label{fig1}
\end{figure}

In this Letter, we study the tunneling properties of ultracold fermions in optical lattices with high accuracy. This is done for pure spin-polarized fermions and for attractive mixtures of fermionic Potassium and bosonic Rubidium atoms. As a key result we directly observe an interaction-induced change in the effective lattice depth and thus a reduction of the fermionic tunneling energy $J$ due to the bosonic component \cite{footnote1}. We explain this in an effective potential picture where mutual self-trapping of both species is expected \cite{Luehmann2008selftrapping, Best2008}. For these measurements we apply a novel multi-band spectroscopy for fermions in optical lattices based on lattice modulation and a subsequent band mapping. Similar to ARPES in solid state physics, our method can accurately resolve the excitation spectrum with full momentum resolution. This allows for the determination of the lattice depth and the tunneling energy with high accuracy but is also promising to study excitons as well as non-equilibrium dynamics of particle and hole excitations.

We prepare a spin-polarized ultracold mixture of bosonic $^{87}$Rb and fermionic $^{40}$K atoms in a crossed optical dipole trap with typically $10^5$ fermions at a temperature below $0.2\,\text{T}_\text{F}$ and $2 \times 10^5$ atoms in the BEC. The dipole trap is operated near the magic wavelength at $\SI{811}{\nano\metre}$, compensating the differential gravitational sag, with an average trap frequency of $\bar{\omega}= 2\pi\times\SI{50}{\hertz}$. We superimpose a 3d optical lattice at $\lambda=\SI{1030}{\nano\metre}$ which we ramp up in $\SI{100}{\milli\second}$. Note that the Rubidium atoms experience an approximately $2.5$ times stronger lattice than the Potassium atoms with respect to their recoil energies due to the different masses and detunings. In the following the lattice strength will be denoted in fermionic recoil energies $\text{E}_\text{r} = \hbar^2 k_\text{L}^2/2m$ with the mass $m$ of $^{40}$K and the lattice vector $k_\text{L}=2\pi/\lambda$. Depending on the lattice depth, the fermionic ground state is a metal or band insulator \cite{Koehl2005,Schneider2008} while the bosons form a superfluid or Mott insulator.

\begin{figure}[t]
  \centering
  \includegraphics[width=\columnwidth]{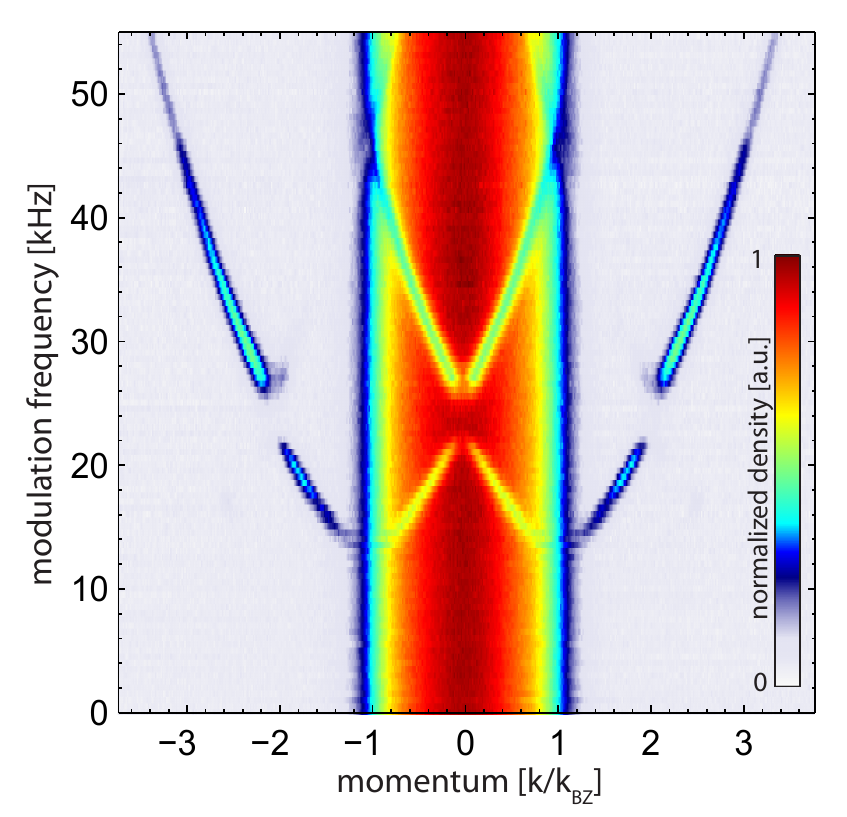}
  \caption{(color online). Momentum-resolved band structure of non-interacting fermions in an optical lattice of $5\,\text{E}_\text{r}$: Shown are the column densities of the momentum distributions for different modulation frequencies. Central plateau are atoms in the first Brillouin zone. Missing particles are holes, representing the reduced zone scheme. Narrow peaks at higher momenta are the outcoupled particles, representing the extended zone scheme.}
  \label{fig2}
\end{figure}

The spectroscopy is performed as follows (see Figure~\ref{fig1}): We excite the system to higher bands by modulating the depth of the optical lattice in one direction for $\SI{1}{\milli\second}$ with a variable frequency and a typical modulation depth of $\SI{15}{\percent}$ \cite{Denschlag2002,Stoeferle2004}. Due to the different curvature of the individual bands, only a specific quasimomentum class of atoms can be resonantly excited at a fixed modulation frequency. Finally we map the quasimomentum distribution onto real momenta by ramping down the optical lattice nonadiabatically in $\SI{200}{\micro\second}$ suppressing band transitions \cite{Greiner2001}. The ramp is designed to be much faster than the trap dynamics to prevent redistribution within each band. Due to the finite ramp time, however, the distribution is smoothed out at the zone boundaries \cite{McKay2009} [see Figure~\ref{fig1}(c)]. We investigate the resulting momentum distribution after a time-of-flight of typically $\SI{20}{\milli\second}$ thus spatially separating different bands and momenta. Applying off-resonant modulation, the atoms stay in the ground state and occupy the first Brillouin zone [see Figure~\ref{fig1}(c)]. If the modulation frequency corresponds to a transition energy, a specific momentum class is excited to a higher band and mapped onto its corresponding Brillouin zone creating a hole within the first Brillouin zone [see Figure~\ref{fig1}(d)]. Note that in contrast to localized particle-hole states in the Mott insulating regime \cite{Joerdens2008} we create particle-hole excitations in momentum space. Varying the modulation frequency gives access to the full band structure with full momentum resolution.

To demonstrate the capabilities of this method, we first apply it to the well-known case of a pure non-interacting fermionic sample in an optical lattice as shown in Figure~\ref{fig2}. Experimentally, we measure the energy difference between the first band and the excited bands, obtaining all information to deduce the full band structure. The outcoupled particles correspond to the extended zone scheme of the band structure. The rise of the band gap and the corresponding flattening of the dispersion are clearly visible at the edge of the Brillouin zones. The holes represent the quasimomenta of the outcoupled atoms and thus the band structure within the reduced zone scheme in a textbook-like manner. Note the inverted curvature of the second band, which becomes apparent in the reversed dispersion of the holes. To our knowledge, this is the first experimental observation of the full band structure for ultracold fermions in optical lattices.

\begin{figure}[t]
  \centering
  \includegraphics[width=\columnwidth]{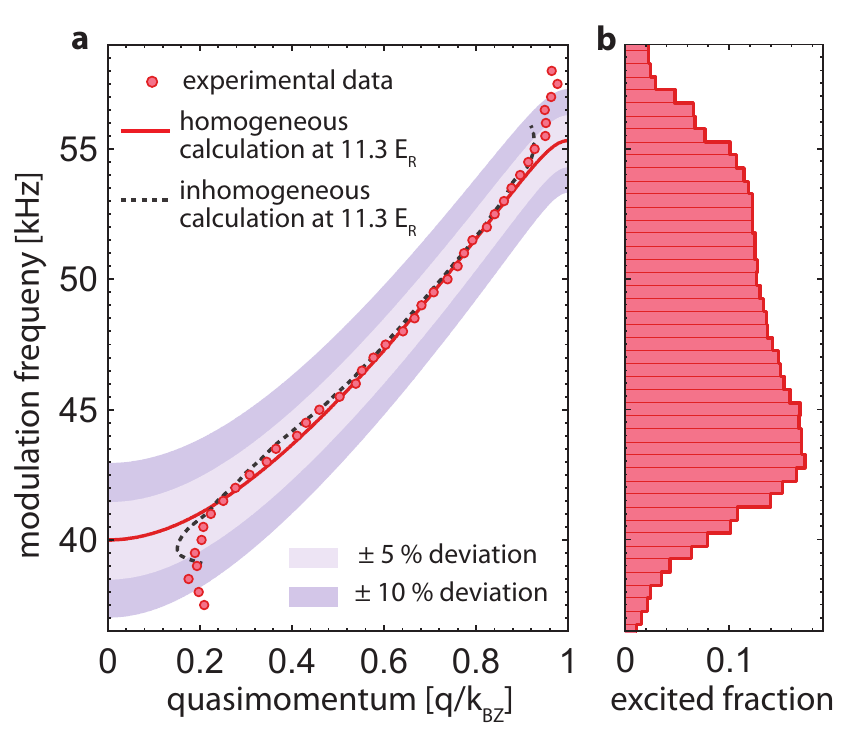}
  \caption{(color online). (a) Extracted dispersion of the third band. Red circles are the centers of mass of the excited atoms. The dashed line is an inhomogeneous 1d calculation, reproducing the data qualitatively. The red solid line is a fit to the data for $0.55<q<0.85\,\text{k}_\text{BZ}$ with a homogeneous 1d band structure. Shaded areas show the dispersion for slightly different lattice depths. (b) Histogram showing the relative number of outcoupled atoms.}
  \label{fig3}
\end{figure}

To analyze the band structure quantitatively, we determine the center of mass of the outcoupled atoms in the time-of-flight pictures and extract the corresponding momenta for all applied modulation frequencies. The result is shown exemplarily in Figure~\ref{fig3}(a) for the third band at a lattice depth of $11\,\text{E}_\text{r}$ with the fermions forming a pure band insulator. The momentum transfer is given in units of  $\text{k}_\text{BZ} = 2\pi/\lambda$. We estimate the lattice depth of the system by a least square fit of a homogeneous 1d band structure (red solid line) to the experimental data. In our data, we clearly observe a deviation of the excitation spectrum from the homogeneous case particularly at small and large momenta. We find good qualitative agreement of this observation comparing our data to a numerical linear response calculation of a finite 1d system including the harmonic confinement. Especially the bending of the spectroscopic signal at the zone edges is well reproduced. In an intuitive picture we interpret this effect as the coupling of different momentum components due to the inhomogeneity especially in the first band. This coupling influences the eigenstates and the density of states at the upper and lower edge of the first band. In these parts of the spectrum no atoms are present to be transferred into the third band. For a specific range of momenta the effect of the inhomogeneity is negligible. Therefore we restrict the fitting to our data to a momentum range between $0.55$ and $0.85\,\text{k}_\text{BZ}$. From this fit we can determine the lattice depth with very high accuracy. The dominant errors of about $2\%$ are of systematic nature, given by the spatial dependence of the lattice depth due to the gaussian shape of our lattice beams. For non-interacting particles, the lattice depth fully determines the tunneling energy. The fitted value of $11.3(2)\,\text{E}_\text{r}$ corresponds to $J/h=\SI{67(3)}{\hertz}$. In addition to the dispersion relation we count the number of transferred atoms in the corresponding Brillouin zone, shown as a histogram in Figure~\ref{fig3}(b). It has a maximum at low momenta and is small where the signal bends away from the homogeneous dispersion as expected.

\begin{figure}[t]
  \centering
  \includegraphics[width=\columnwidth]{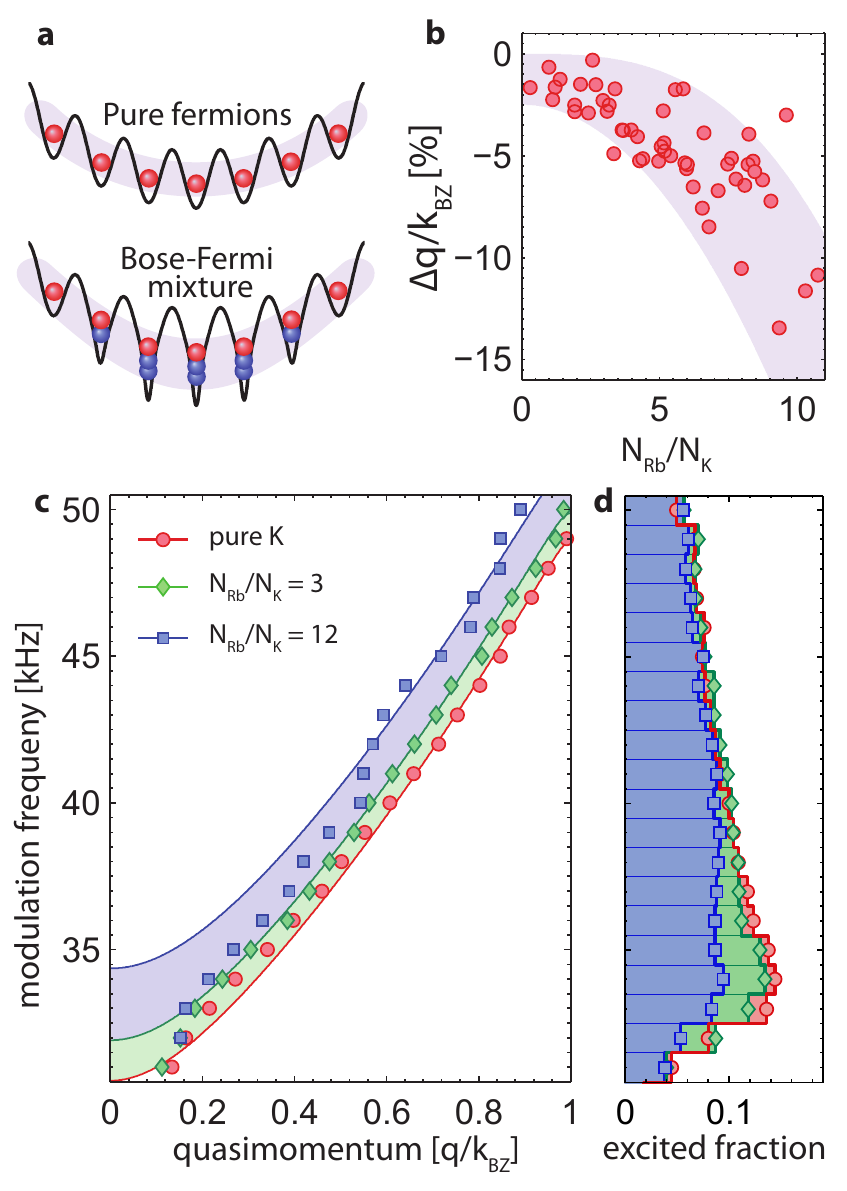}
  \caption{(color online). (a) Sketch of the effective potential for different bosonic fillings. Fermions are more tightly bound if attractive bosons are added. (b) Shift of the outcoupled momentum against the relative atom number $N_\text{Rb}/N_\text{K}$ at $5\,\text{E}_\text{r}$ and fixed modulation frequency of $\SI{34}{\kilo\hertz}$. Lower outcoupled momenta correspond to a deeper lattice. The shaded area is a guide to the eye. (c) Fermionic band structure for different bosonic admixtures. Solid lines are fitted dispersions. (d) Histogram showing the relative number of outcoupled atoms.}
  \label{fig4}
\end{figure}

We now turn to our main results, investigating the influence of interspecies interaction on the fermionic band structure: We apply our spectroscopy method to study an attractive mixture of bosonic $^{87}$Rb atoms in the state $\left|1,1\right\rangle$ and fermionic $^{40}$K atoms in the state $\left|9/2,9/2\right\rangle$. In this system, interaction effects represented by the strong attractive interspecies scattering length $\text{a}_\text{bf} \approx -215\,\text{a}_\text{0}$ between $^{40}$K and $^{87}$Rb \cite{Ferlaino2006} and the repulsive bosonic scattering length $\text{a}_\text{bb}\approx 100\,\text{a}_\text{0}$, play a crucial role. We choose lattice depths such that the Rubidium atoms form a Mott insulator and the Potassium atoms a band insulator. Due to the $2.5$ times higher lattice depth the tunneling energy of the bosons is very small, while the fermions experience a moderate lattice potential with high tunneling energies. We describe this situation in an effective potential picture \cite{Best2008,Luehmann2008selftrapping}: If a boson and a fermion are at the same lattice site, the attractive interaction leads to an enhanced localization and thus a suppression of the tunneling. This can be described as an effective potential which is proportional to the respective density of the other component. Since both species mutually influence each other, this is called self-trapping. The local deformation of the wavefunctions can be interpreted as the admixture of higher bands. The resulting potential for the fermionic atoms due to the admixture of bosons is sketched in Figure~\ref{fig4}(a). In a homogeneous system with a fixed number of Rubidium atoms per site the effective potential for the fermions has the same periodicity as the original lattice but the effective lattice depth is significantly increased. Here, for the first time we are able to directly observe this interaction shift for different mixture ratios of Potassium and Rubidium. To maximize the signal-to-noise ratio, we restrict our measurements to the third band, which has the highest excitation rate.
 
A detailed scan of the third band is shown in Figure~\ref{fig4}(c). The purely fermionic system (red circles) at a lattice depth of $7.6\,\text{E}_\text{r}$ shows the typical characteristics at low momenta, that can be qualitatively understood as in Figure~\ref{fig3}. The admixture of bosons leads to a clear shift of the band structure and a suppression of the excitation fraction. For moderate atom numbers with a bosonic Mott insulator of mainly unity filling, there is a small but distinct overall shift of the spectroscopic signal (green diamonds). Using the same fit as for the pure fermionic case, we obtain a lattice depth of $8,2\,\text{E}_\text{r}$. This is consistent with the theoretical prediction without correlation effects. For higher particle numbers a shell structure for the bosons emerges. Here we estimate a maximal occupation of two in the Mott insulator. In this case, the spectroscopic signal shows large deviations from the simple model (blue squares). We attribute this to the increased inhomogeneity of the effective potential due to the bosonic shell structure. If we fit a band structure as before we obtain a lattice depth of $9.2\,\text{E}_\text{r}$ which is significantly higher than the uncorrelated shift for two bosons per site. Assuming the fermions as still non-interacting, we can derive their tunneling energy from the lattice depth with a simple calculation. We obtain values for $J/h$ of $\SI{160}{\hertz}$, $\SI{138}{\hertz}$, and $\SI{109}{\hertz}$ for the three respective cases showing a clear reduction of the fermionic tunneling due to the admixture of bosons. The number of excited atoms is also consistent with our interpretation: Since the fermions experience a higher lattice depth in the presence of bosons, less atoms are transferred at low energies, as clearly visible in Figure~\ref{fig4}(d).

To investigate the particle-number dependence in more detail, we performed a measurement at fixed modulation frequency for a shallow lattice of $5\,\text{E}_\text{r}$ [see Figure~\ref{fig4}(b)]. For pure fermions, the modulation frequency of $\SI{34}{\kilo\hertz}$ results in a momentum of $k = 0.58\,\text{k}_\text{BZ}$. For an increasing admixture of bosons the mean outcoupled momentum decreases. This is consistent with the effective potential approach which predicts an increase of the effective lattice depth as in Figure~\ref{fig4}(c). The non-linear slope indicates enhanced localization due to stronger correlation effects for higher particle numbers.

In conclusion, we have performed a fully momentum resolved spectroscopy of ultracold fermions in optical lattices for the first time. We have investigated a mixture of fermions and localized bosons in an optical lattice and observed an interaction induced shift of the fermionic band structure due to the attractive bosons as expected from an effective potential approach. The effective lattice depth was increased up to $20\%$ while the corresponding tunneling energy was reduced by more than $30\%$. Further we found evidence for a correlation induced enhancement of the interaction shift. As a complementary measurment to \cite{Will2010}, this is an important step for the comparison with theory. In agreement with \cite{Best2008,Will2010}, we conclude that a simple single-band Hubbard model is not sufficient for the realistic description of strongly attractive quantum gas mixtures. Moreover, we observe a confinement-induced deformation of the spectroscopy signal at small momenta, consistent with numerical calculations. Our easy-to-implement spectroscopy method is also promising for the investigation of interacting fermi gases in optical lattices. It is in particular suited for band gap excitations such as excitons and the dynamics of particle and hole excitations.

We thank E.~Demler, D.~Pfannkuche, and F.~Zhou for valuable discussion. We acknowledge financial support by DFG via grant FOR801.

\cleardoublepage


\begin{thebibliography}{50}
\bibitem{footnote1}
The tunneling energy $J$ is defined in tight-binding as $J=\int \psi_{i+1}^*(x)[\tfrac{p^2}{2m}+V\sin(k_\text{L}x)^2]\psi_{i}(x)dx$ with the local Wannier functions $\psi_i(x)$ at site $i$ and the lattice depth $V$.
\end{thebibliography}
\end{document}